\documentclass[10pt]{article}
\usepackage[utf8]{inputenc}
\usepackage{enumerate}
\usepackage[utf8]{inputenc}
\usepackage{amsmath}
\usepackage{amsfonts}
\usepackage{amssymb}
\usepackage{graphicx}
\usepackage{color}
\usepackage{mathptmx}
\usepackage{multicol}
\usepackage{float}
\RequirePackage[colorlinks=true,citecolor=blue,urlcolor=blue,linkcolor=blue]{hyperref}
\usepackage[left=2cm,right=2cm,top=2cm,bottom=2cm]{geometry}

\author{Tanima Duary\footnote{Email: td14ip021@iiserkol.ac.in} ~ and  Narayan Banerjee\footnote{Email: narayan@iiserkol.ac.in} \\ {\normalsize Department of Physical Sciences, 
Indian Institute of Science Education and Research Kolkata,
Mohanpur - 741 246, WB, India}}
\title{{\Large\bf Brans-Dicke Cosmology: Thermodynamic viability}}
\begin{document}
\date{}
\maketitle
\vspace{-0.5cm}
\begin{abstract}
Brans-Dicke cosmological models for a spatially isotropic and homogeneous universe are tested in terms of the validity of the Generalized Second Law of Thermodynamics (GSL). The investigation is carried out in the Einstein frame. It is found that the models are thermodynamically viable for negative values of the Brans-Dicke parameter $\omega$ and thus are quite consistent with the recent accelerated expansion of the universe.
\end{abstract}


\medskip

\section{Introduction}

It is quite well known that Brans-Dicke theory (BDT) of gravity\cite{BD61} often becomes a point of discussion for its potential in resolving issues related to cosmology. Two worthwhile examples are the extended inflation\cite{johri, la-stein} that resolves the graceful exit problem of the inflationary scenario and the possibility of a late time acceleration of the universe even without a dark energy\cite{NBDP}. BDT has its limitations. One is the fact that the characteristic coupling constant of the theory, $\omega$, has to pick up a very high value, for compatibility with local astronomical tests\cite{will-new}, whereas for a successful resolution of the cosmological problems, $\omega$ requires a small value. Another is that the said reduction of BDT to GR in the infinite $\omega$ limit has been proved to be true only for a limited situation\cite{soma, valerio}. In spite of all these, the theory still enjoys the attention for its applicability in many real cosmological problems and also for its formal similarity with many nonminimally coupled gravity theory. \\

Thermodynamics is surely amongst the most robust branches of science, as it has remained unaffected by many new discoveries and new concepts. The motivation of this note is to check the thermodynamic viability of cosmological models in BDT. Certainly the principal goal is to check this viability in the matter dominated regime of the universe when the universe has a transition from a decelerated to an accelerated phase of expansion. However, the radiation dominated regime is also taken care of, for the sake of completeness. Thermodynamic analysis for cosmological models was also done by Bhattacharya and Debnath\cite{ujjwal} in a different context. They considered an accelerated expansion in an extended BDT by including a potential and the BD parameter $\omega$ was considered to be a function of the scalar field. The present investigation considers a pure BDT, which can drive the accelerated expansion by its own right.\\

For the sake of tractability, we consider the equations in the so called Einstein frame achieved by a suitable conformal transformation\cite{dicke62}. This naturally leads to a situation where the matter distribution does not evolve independent of the BD scalar field. So the Bianchi identity will lead to a conservation of the BD field and the matter distribution taken together. This feature is consistent with the speculation that there is no apriori reason to believe that the dark matter and dark energy evolve independently without any nongravitational interaction amongst each other\cite{fernando}. There are already some investigations where the BD scalar field is allowed to interact with matter\cite{sudipta1, sudipta2}. \\

The findings of this investigation is quite encouraging. The models, particularly for the universe filled with a pressureless fluid, is consistent with only negative values of the parameter $\omega$, and that too in the range for which BDT can drive the accelerated expansion by itself. \\

In the next section we write down the Einstein-BD field equations in Einstein frame. In section 3 the thermodynamic viability is considered for the radiation dominated and matter dominated eras and the 4th and final section includes some concluding remarks.

\section{Brans-Dicke Theory in Einstein frame}

\hspace{3mm}

Brans-Dicke theory\cite{BD61} is given by the action
\begin{equation}\label{action1}
 \mathcal{A} = \frac{1}{16\pi G_0}\int\sqrt{-g}\left[\phi R-\omega\frac{\phi_{,\alpha}\phi^{,\alpha}}{\phi}+\mathcal{L}_M\right] d^4x ,
\end{equation}

where $\phi$ is the scalar field, nonmiminally coupled to gravity, $\omega$ is a dimensionless constant, $R$ is the Ricci scalar and $\mathcal{L}_M$ is the Lagrangian density due to the matter ditribution. With a conformal transformation\cite{dicke62} 

\begin{equation}\label{conformal}
 \bar{g}_{\mu\nu} = \phi g_{\mu\nu},
\end{equation}

the action becomes 
\begin{equation}\label{action2}
 \bar{\mathcal{A}} =\frac{1}{16\pi G_0} \int \sqrt{-\bar{g}}\left[\phi_0\left(\bar{R}-\frac{2\omega+3}{2}\psi_{,\alpha}\psi_{,\beta}\bar{g}^{\alpha\beta}\right)+\bar{\mathcal{L}_M}\right]d^4x ,
\end{equation}

where $\psi = \ln (\frac{\phi}{\phi_0}),$ $\phi_0$ being a constant.

In this revised version, popularly called as the ``Einstein frame'', the field equations look much simpler, 

\begin{align}\label{B1}
G_{\alpha\beta} = T_{\alpha\beta}+\frac{2\omega+3}{2}(\psi_{,\alpha}\psi_{,\beta}-\frac{1}{2}g_{\alpha\beta}\psi^{,\mu}\psi_{,\mu}).
\end{align}
This is written in the units such that $8\pi G_0=1$.
The equation of motion for the scalar field $\psi$, obtained by varying the action (\ref{action2}) by $\psi$, is

\begin{align}\label{B2}
\Box \psi =\frac{T}{2\omega+3},
\end{align}

$T$ being the trace of the energy momentum tensor for the matter sector. For a matter sector given by a perfect fluid, the energy momentum tensor looks like 

\begin{align}\label{B3}
T_{\alpha\beta}= (\rho+p)v_\alpha v_\beta +p g_{\alpha\beta},
\end{align}

where $\rho , p$ are the energy density and pressure of the fluid and $v_{\alpha}$ is the unit timelike vector,  $v^\mu v_\mu =-1$. In a comoving coordinate system, $v^\mu =\delta^\mu_0$.  It is to be noted that $\rho$ and $p$ are in revised unit, we have avoided the ``bar" sign here. For a spatially flat, homogeneous and isotropic cosmology given by the metric

\begin{equation}\label{metric}
\mathrm{d}s^2 = -\mathrm{d}t^2+{a(t)}^2 [\mathrm{d}x^2+\mathrm{d}y^2+\mathrm{d}z^2],
\end{equation}

where $a$ is the scale factor, the Einstein-Brans-Dicke field equations in the conformally transformed version\cite{dicke62} take the form, 

\begin{align}\label{B4}
 3\left(\frac{\dot{a}}{a}\right)^2 = \rho+\rho_\psi ,
\end{align}
\begin{align}\label{B5}
 2\left(\frac{\ddot{a}}{a}\right)+\left(\frac{\dot{a}}{a}\right)^2 = -p-p_\psi, 
\end{align}

where $\rho_\psi = p_\psi= \frac{2\omega+3}{4}\dot{\psi}^2$ denotes the energy density and pressure of the scalar field. The wave equation of the scalar field takes the form, 
 \begin{align}\label{B6}
 \ddot{\psi}+3 \frac{\dot{a}}{a}\dot{\psi}=\frac{-T}{2\omega+3}.
 \end{align}

In this revised version, the equations formally resemble those for two component matter, one is the fluid and the other being a massless scalar field $\psi$.  The continuity equation for the fluid and scalar field taken together, looks like \begin{align}\label{B7}
 \dot{\rho}_t= -3\frac{\dot{a}}{a}(\rho_t+p_t),
\end{align}
where suffix $t$ denotes the total quantity and is the sum of the fluid part and the scalar field part. This equation is not an independent one, as it follows from the Bianchi identities.\\

The kinematical quantities of interest for the present work are the Hubble parameter defined as, $H = \frac{\dot{a}}{a}$ and the deceleration parameter defined as, $q=-\frac{\ddot{a}/a}{\left(\dot{a}/a\right)^2}.$

\section{The Thermodynamic Consideration:}
 
 According to the Generalised Second Law of Thermodynamics (GSL), total entropy of the Universe ($ S_{tot}$) can not decrease with time\cite{Bekenstein1972,Bekenstein1973,Bekenstein1974}. The total entropy is sum of horizon entropy ($S_h$) and the fluid (inside the horizon) entropy ($S_{in}$) i.e.,
\begin{equation} \label{T1}
S_{tot}=S_h+S_{in}.
\end{equation}
 For an expanding universe, it is more relevant to consider the entropy of dynamical apparent horizon rather than teleological event horizon. Entropy of the apparent horizon is given by the relation\cite{Bak2000},
\begin{equation}\label{T2}
S_h=2\pi A,
\end{equation}
where $A$ denotes the area of the apparent horizon and is related to radius of the apparent horizon ($R_h$) as, $A = 4\pi R_h^2$.
In flat FRW spacetime, the horizon radius $R_h$ is related to Hubble parameter\cite{Bak2000} as,
\begin{equation}\label{T3}
R_h = \frac{1}{H}.
\end{equation}

Therefore, rate of change in horizon entropy is 
\begin{align} \label{T4}
\dot{S}_h =-16\pi^2\frac{\dot{H}}{H^3}.
\end{align}

Applying Gibb's law of thermodynamics for the fluid inside the horizon we get, 
\begin{equation} \label{T5}
T_{in}\mathrm{d}S_{in} = \mathrm{d}E_{in}+p_t\mathrm{d}V_h.
\end{equation}
The rate of change in entropy of the fluid inside the horizon can now be written as,
\begin{equation}\label{T6}
\dot{S}_{in}=\frac{1}{T_{in}} [(\rho_t+p_t)\dot{V}_h+\dot{\rho}_t V_h],
\end{equation}
where the volume $V_h=\frac{4}{3}\pi R_h^3$. If we consider that the fluid is in thermal equilibrium with the horizon, then $T_{in}$ is same as the temperature of the dynamical apparent horizon($T_h$) i.e., Hayward-Kodama temperature\cite{hayward1998, hayward2009, book:Faraoni}, which is expressed as,
\begin{equation} \label{T7}
T_{h} = \frac{2H^2+\dot{H}}{4\pi H}.
\end{equation}
It could be seen that in de Sitter space(for which $\dot{H}=0$), this temperature reduces to the Hawking temperature, $T_{\text{Hawking}}=\frac{H}{2\pi}$ \cite{Hawking1974}. 

Using Eq.\eqref{T7} and the field equations in \eqref{T6}, rate of change of entropy inside the horizon is obtained as,

\begin{equation}\label{T9}
\dot{S}_{in} =16\pi^2 \frac{\dot{H}}{H^3}\left(1+\frac{\dot{H}}{2H^2+\dot{H}}\right).
\end{equation}

Adding Eq.\eqref{T4}and \eqref{T9}, we get the rate of change in total entropy as,
\begin{equation}\label{T10}
\dot{S}_{tot} =16\pi^2 \frac{\dot{H}^2}{H^3}\left(\frac{1}{2H^2+\dot{H}}\right).
\end{equation}

\subsection{Radiation Era :}
\hspace{3mm} 
In case of radiation, the equation of state of the fluid is, $p=\frac{1}{3}\rho$. Hence, the trace of the stress-energy tensor vanishes. Consequently, the  wave equation \eqref{B6} can easily be integrated to yield the following relation, 
\begin{align}\label{r1}
\dot{\psi}= \frac{\alpha}{a^3},
\end{align}
where $\alpha$ is an integration constant.

From the field equations \eqref{B4}-\eqref{B5}, using the equation of state and equation \eqref{r1}, one can write,
\begin{align}\label{r2}
\ddot{a}+ \frac{\dot{a}^2}{a}=-\frac{\beta}{a^5}, 
\end{align}
 where $\beta$ is a constant, given as $\frac{2\omega+3}{12}\alpha^2$. 
 
 Integration of the above equation results in 
 \begin{align}\label{r3}
 \dot{a}^2= \frac{\sigma a^2+\beta}{a^4},
 \end{align}
where $\sigma$ is the integration constant.
The nature of the solutions of this equation differs depending on the sign of $\sigma$ and $\beta$. One should note from equation \eqref{r3} that both $\sigma$ and $\beta$ cannot be negative simultaneously.

\subsubsection{\underline{Case-I : $\sigma>0$, $\beta>0$}}  
\hspace{3mm} In this case, the solution of wave eqn \eqref{B6} is given by, 
\begin{align}\label{r4}
\psi+\psi_0 = \frac{\alpha}{2\sqrt{\beta}}\ln\left| \frac{\sqrt{a^2+\beta/\sigma}-\sqrt{\beta/\sigma}}{\sqrt{a^2+\beta/\sigma}+\sqrt{\beta/\sigma}} \right|.
\end{align}

Integrating eqn \eqref{r3}, we obtain the relation between the scale factor and time as,
\begin{align}\label{r5}
t+t_0= \frac{1}{2\sqrt{\sigma}}a\sqrt{a^2+\beta/\sigma}-\frac{1}{2}\frac{\beta}{\sigma^{3/2}}\ln \left| a+\sqrt{a^2+\beta/\sigma}\right|.
\end{align}
Throughout this article, suffix zero implies value of the quantity at present epoch.
This set of solutions for the scale factor is already known\cite{ABanerjee1985, NB1985}.
Substituting $H$ and $\ddot{a}$ in the definition of deceleration parameter, we get
\begin{align}\label{r6}
q = 1+\frac{\beta}{\beta+\sigma a^2} >0.
\end{align}
Therefore this model leads to an ever-decelerating universe.
\begin{itemize}
\item \textit{\textbf{\underline{Verification of GSL}:}}
\end{itemize}
\hspace{3mm} 
From eqn \eqref{T10}, we obtain the rate of change in total entropy as, 
\begin{align}\label{r8}
\dot{S}_{tot}=-64\pi ^2 \frac{\sqrt{\sigma}}{\beta}\frac{a^3\left(a^2+3\beta/2\sigma\right)^2}{\left(a^2+\beta/\sigma\right)^{3/2}}.
\end{align}
Since $\beta$ and $\sigma$ both are positive, it is clear that $\dot{S}_{tot}<0$. This case therefore clearly fails the GSL test.

\subsubsection{\underline{Case-II : $\sigma>0$, $\beta<0$}}  
\hspace{3mm} $\beta$ is negative if $\omega<-3/2$. We write $\beta=-\gamma^2$, where $\gamma$ is real. 
Hence eqn \eqref{r3} turns into, 
\begin{align}\label{r9}
\dot{a}^2=\frac{\sigma a^2-\gamma^2}{a^4}.
\end{align}
The above equation suggests that there is a bounce at $a= \gamma/\sqrt{\sigma}$. Furthermore, at the point of bounce, the proper volume ($a^3$)  should be a minimum as $\ddot{a} = - \frac{\beta}{a^5} > 0$. One can also check that the total density, pressure also remain finite. So this bounce apparently avoids the Big Bang singularity. However, there is a caveat, at the point bounce, one has $a= \gamma/\sqrt{\sigma}$, but lower values of $a$ is not allowed by the equation \eqref{r9}. So there is clearly a discontinuity. \\

In this case, we get the solution of wave eqn \eqref{B6} as, 
 
\begin{align}\label{r10}
\psi+\psi_0 = \frac{\alpha}{\gamma}\arctan\left(\sqrt{\frac{\sigma a^2}{\gamma^2}-1}\right),
\end{align}

and the solution for the scale factor as 
\begin{align}\label{r11}
t+t_0 = \frac{1}{2\sqrt{\sigma}}a\sqrt{a^2-\gamma^2/\sigma}+\frac{1}{2}\frac{\gamma^2}{\sigma^{3/2}}\ln \left|a+\sqrt{a^2-\gamma^2/\sigma}\right|. 
\end{align}
Accordingly, the deceleration parameter can be obtained as, 
\begin{align}\label{r12}
q=-\frac{2\gamma^2/\sigma-a^2}{a^2-\gamma^2/\sigma}.
\end{align}

Therefore, we get an accelerated phase in the region of $\gamma^2/\sigma<a^2<2\gamma^2/\sigma$. When $a^2>2\gamma^2/\sigma$, the universe has a decelerated expansion. It should be noted that the model does not work for $a^2 < \gamma^2/\sigma$.

\begin{itemize}
\item \textit{\textbf{\underline{Verification of GSL}:}}
\end{itemize}
\hspace{3mm} 

From eqn \eqref{T10} and \eqref{r9}, we get, 

\begin{align}\label{r13}
\dot{S}_{tot}=64\pi ^2 \frac{\sqrt{\sigma}}{\gamma^2}\frac{a^3\left(a^2-3\gamma^2/2\sigma\right)^2}{\left(a^2-\gamma^2/\sigma\right)^{3/2}}.
\end{align}
Since $a\nless\gamma/\sqrt{\sigma}$, the rate of change of total entropy is positive definite. Hence, GSL is satisfied in both of the accelerated and decelerated phases.

\subsubsection{\underline{Case-III : $\sigma<0$, $\beta>0$}}  
\hspace{3mm}

Let us consider $\sigma=-\lambda^2$. Accordingly, eqn\eqref{r3} modifies as, 
\begin{equation}\label{r14}
\dot{a}^2= \frac{\beta-\lambda^2a^2}{a^4}.
\end{equation}
The equation governing the variation of scalar-field with scale factor is 
\begin{align}\label{r15}
\psi+\psi_0 = \frac{\alpha}{2\sqrt{\beta}}\ln \left|\frac{\sqrt{\beta/\lambda^2-a^2}-\sqrt{\beta/\lambda^2}}{\sqrt{\beta/\lambda^2-a^2}+\sqrt{\beta/\lambda^2}}\right|,
\end{align}

which is obtained by integrating the wave equation using eqn \eqref{r1}.

Direct integration of eqn\eqref{r14} gives the solution for the scale factor as 
\begin{equation}\label{r16}
t+t_0 =- \frac{1}{2\lambda}a\sqrt{\beta/\lambda^2-a^2}+\frac{\beta}{2\lambda^3}\arcsin\left(\frac{a}{\sqrt{\beta/\lambda^2}}\right).
\end{equation}

From eqn \eqref{r14}, it is obvious that $a\ngtr \sqrt{\beta}/\lambda$. At $a=\sqrt{\beta}/\lambda$,  $\dot{a}$ becomes zero and  $\ddot{a} = -\lambda^5/\beta^{5/2}$. But the model is not defined beyond $\sqrt{\beta}/\lambda$, so this is not really a maximum of the scale factor. In fact the model is not valid beyond that. Since there is no limit in the lower bound of $\ddot{a}$, it will eventually hit the singularity if we trace back in the past. Deceleration parameter is given by, 
\begin{equation}\label{r17}
q  = \frac{2\beta-\lambda^2 a^2}{\beta-\lambda^2 a^2},
\end{equation}
which is positive, since $a\ngtr \sqrt{\beta}/\lambda$. Therefore, this model leads to a deceleration phase. 
\begin{itemize}
\item \textit{\textbf{\underline{Verification of GSL}:}}
\end{itemize}
\hspace{3mm} 
In this case, we get rate of change in total entropy as, 
\begin{equation}\label{r18}
\dot{S}_{tot}= -\frac{64\pi^2 \lambda}{\beta}\frac{a^3\left(3\beta/2\lambda^2-a^2\right)^2}{\left(\beta/\lambda^2-a^2\right)^{3/2}}.
\end{equation}

The rate of change in total entropy is negative, since $a\ngtr \sqrt{\beta}/\lambda$. Hence this models does not comply with GSL.

\subsection{Dust Era :}
\hspace{3mm} 
The next example is that of a pressureless matter ($p=0$). Therefore, the field equations are given as, 
\begin{align}\label{D1}
 3\left(\frac{\dot{a}}{a}\right)^2 = \rho+\rho_\psi ,
\end{align}
\begin{align}\label{D2}
2\left(\frac{\ddot{a}}{a}\right)+\left(\frac{\dot{a}}{a}\right)^2 = -p_\psi. 
\end{align}

Putting $p=0$ in the wave eqn \eqref{B6} we get,
\begin{equation}\label{D3}
 \ddot{\psi}+3 \frac{\dot{a}}{a}\dot{\psi}=\frac{\rho}{2\omega+3}.
\end{equation}
Differentiating the field Eq. \eqref{D1} and then using the field Eqs. \eqref{D1}-\eqref{D2} and wave eqn \eqref{D3} , we get the modified version of continuity eqn as, 
\begin{equation}\label{D4}
\dot{\rho}+3\frac{\dot{a}}{a} \rho = -\frac{\rho\dot{\psi}}{2}.
\end{equation}
The non-zero term in right hand side of the above eqn implies that the energy densities of two fluids, viz, dust and scalar field, are not conserved individually. But they satisfy the total conservation eqn, 
\begin{align}\label{D5}
 \dot{\rho}_t= -3\frac{\dot{a}}{a}(\rho_t+p_t),
 \end{align}
 which can easily be obtained from the field equations. 
 Integrating eqn \eqref{D4} we get, 
 \begin{equation}\label{D6}
 \rho= \rho_0 \frac{\exp[-\frac{\psi}{2}]}{a^3}.
 \end{equation}
Using the wave equation for eliminating $\rho$, an integration of a combination of the field equations yields, 
\begin{equation}\label{D7}
a^2\dot{a} = \frac{2\omega+3}{2}a^3\dot{\psi}+\xi, 
\end{equation}

where $\xi$ is an integration constant. 
Substituting $\dot{\psi}$ from the above equation into Eq.\eqref{D2},  we obtain,
\begin{equation}\label{D8}
2\ddot{a}=-\chi\frac{\dot{a}^2}{a}+\frac{2\xi}{2\omega+3}\frac{\dot{a}}{a^3}-\frac{\xi^2}{2\omega+3}\frac{1}{a^5},
\end{equation}
where $\chi=\frac{2\omega+4}{2\omega+3}$ is a constant related to Brans-Dicke parameter $\omega$. 
In order to proceed further, we take the constant of integration $\xi=0$ without loss of generality.
Hence Eq.\eqref{D8} becomes, 
\begin{equation}\label{D9}
2\ddot{a}=-\chi\frac{\dot{a}^2}{a}.
\end{equation}
Now the equation is easily integrable and we get 
\begin{equation}\label{D10}
\dot{a}= \mu a^{-\chi/2},
\end{equation}
where $\mu$ is an integration constant.  
 
From Eq.\eqref{D7} we obtain, 

\begin{equation}\label{D11}
\dot{\psi}= \frac{2\mu}{2\omega+3}a^{-(\frac{\chi}{2}+1)}.
\end{equation}

Integration of the above two equations results in the solution for the system as, 
\begin{align}\label{D12}
t+t_0 &=\frac{1}{\mu(\frac{\chi}{2}+1)}a^{\frac{\chi}{2}+1}, \\
\psi+\psi_0 &= \ln\left(a^{\frac{2}{2\omega+3}}\right),
\end{align}
where $t_0$ and $\psi_0$ are constants of integration.
Now from the definition, we obtain deceleration parameter as, 
\begin{equation}\label{D13}
q = \frac{\omega+2}{2\omega+3}.
\end{equation}

From eqn \eqref{D12} it is clear that, $\mu$ and $\frac{\chi}{2}+1$ both have to be positive in order to model an expanding universe. The positivity of $\frac{\chi}{2}+1$ can be achieved in two ways, and we discuss them separately. 

\subsubsection{\underline{Case-I :}}  
\hspace{3mm}
$\frac{\chi}{2}+1 = \frac{3\omega+5}{2\omega+3}>0$. In the first case, we consider both $3\omega+5$ and $2\omega+3$ are positive, which corresponds $\omega>-3/2$. Consequently we get expansion of the universe to be decelerating.
\begin{itemize}

\item \textit{\textbf{\underline{Verification of GSL}:}}
\end{itemize}
\hspace{3mm} 
From eqn \eqref{T10} we get the rate of change in total entropy as, 
\begin{equation}\label{D14}
\dot{S}_{tot} = 16\pi^2\left(\frac{\chi}{2}+1\right)^3\left(\frac{2}{2-\chi}\right)t.
\end{equation}
 For  $\dot{S}_{tot}$ to be positive,  $\omega$ has to be greater than $-1$. 

\subsubsection{\underline{Case-II :}}  
\hspace{3mm}
 
An alternative way to make $\frac{\chi}{2}+1>0$ is that both $3\omega+5$ and $2\omega+3$ are negative, which corresponds $\omega<-5/3$. In this case one can achieve cosmic acceleration for $-2<\omega<-5/3$. However, if $\omega<-2$, one has a decelerated expansion for the univere (from the equation (\ref{D13})). The bound on the values on $\omega$ is consistent with that obtained by Banerjee and Pavon\cite{NBDP}.

\begin{itemize}
\item \textit{\textbf{\underline{Verification of GSL}:}}
\end{itemize}
\hspace{3mm} 
Form eqn \eqref{D14}, it is obvious that for $\dot{S}_{tot}$ to be positive, $\chi$ has to be less than $2$. Since $\omega<-5/3$ in this case,  $\chi$ is surely less than $2$. Consequently, $\dot{S}_{tot}>0$.  Hence, this model satisfies GSL. 

\section{Concluding Remarks :}
\hspace{3mm} 

This work investigates the thermodynamic behaviour of radiation and dust dominated cosmological models for a spatially flat, homogeneous and isotropic universe in Brans-Dicke theory, in its conformally transformed version, popularly called the ``Einstein frame". We focus our discussion on validity of the generalized second law of thermodynamics\cite{Bekenstein1972,Bekenstein1973,Bekenstein1974}, the net entropy of the universe, which is a combination of the matter entropy and the horizon entropy, cannot decrease with time. \\

It is very surprising to see that for a radiation dominated universe, BDT solutions with a positive definite $\omega$ does not respect the GSL. For some ranges of the negative values of $\omega$, however, the model passes the thermodynamic fitness test. This is quite encouraging, as an accelerated expansion of the universe requires a negative value of the parameter $\omega$. \\

For a dust dominated universe, the model does satisfy the GSL for some small negative values of $\omega$. The range, $-2< \omega < -\frac{5}{3}$, has a strong overlap with that required for an accelerated expansion without any exotic matter\cite{NBDP}. \\

The surprise element found in this investigation is that BDT enjoys thermodynamic support for the values of $\omega$ that actually drives the alleged accelerated expansion of the universe, and not for the decelerated expansion in general. \\

{\bf Acknowledgment}: \\

Tanima Duary conveys sincere gratitude to the Council of Scientific and Industrial Research (CSIR), Government of India, for supporting this project financially.  She thanks Aritra Biswas for efficacious suggestions and discussions.

\end{document}